\documentclass[letterpaper, 10 pt, conference]{ieeeconf}
\usepackage{cite}

\usepackage{amsmath,amssymb,amsfonts}

\usepackage{graphicx}
\usepackage{textcomp}
\usepackage{xcolor}

\usepackage{amsmath,amsthm,amssymb,latexsym}
\usepackage{array,multirow,alltt,enumerate,bbm,eucal,graphicx}
\usepackage[colorlinks]{hyperref}
\usepackage{mathpazo,xcolor,stmaryrd,mathpartir}
\usepackage{tikz,pgfmath}
\usepgflibrary{shapes}
\usetikzlibrary{arrows,automata,backgrounds}
\newtheorem{theorem}{Theorem}
\newtheorem{lemma}{Lemma}
\newtheorem{conjecture}{Conjecture}
\newtheorem*{remark}{Remark}
\newtheorem*{definition}{Definition}
\usepackage{cleveref}
\usepackage{algorithm}
\usepackage{algpseudocode}

\usepackage{pseudocode}
\crefname{lemma}{Lemma}{Lemmas}
\IEEEoverridecommandlockouts  
\title{\LARGE \bf Optimizing Queues with Deadlines under Infrequent Monitoring}
\author{Faraz Farahvash and Ao Tang 
\thanks{Faraz Farahvash and Ao Tang are with the School of Electrical and Computer Engineering, Cornell University, Ithaca, NY 14853, USA. Emails: \{\tt\small ff227,atang\}@cornell.edu}
}

\begin{document}

\maketitle
\begin{abstract}
In this paper, we aim to improve the percentage of packets meeting their deadline in discrete-time M/M/1 queues with infrequent monitoring. More specifically, we look into policies that only monitor the system (and subsequently take actions) after a packet arrival. We model the system as an MDP and provide the optimal policy for some special cases. Furthermore, we introduce a heuristic algorithm called "AB-n" for general deadlines. Finally, we provide numerical results demonstrating the desirable performance of "AB-n" policies. 
\end{abstract}
\section{Introduction}
In recent years, there have been a lot of queuing systems where packets have a deadline to meet \cite{Wilson2011}, \cite{Alizadeh2010}, \cite{Vamanan2012}. Such queuing systems are called "real-time queuing systems" in the literature \cite{Kruk2011}. In these systems, packets missing their deadline are discarded. Thus, using smart queuing policies to increase the percentage of packets meeting their deadline is needed. 

The classical result proposed by \cite{Panwar1988} states that the "Earliest Deadline First (EDF)" policy minimizes the percentage of packets missing their deadline in $G/M/c+G$ queues. There have been numerous papers analyzing EDF. An approximation performance analysis is done by \cite{Hong1989}. Also, \cite{Kruk2011} does a heavy traffic analysis for EDF queues. \cite{shakkottai1999} argues that EDF is not necessarily optimal for wireless channels where queue failures can happen. Other related works include but are not limited to \cite{haritsa1991},\cite{zhang2021}, \cite{raviv2018}, \cite{Doytchinov2001}, and \cite{Atar2015}.

A key assumption for the optimality of EDF policies is that a packet is dropped from the queue (or server) as soon as its deadline elapses. Enforcing that requires the policy to constantly monitor the system. This is not always feasible. In this paper, we will look into cases where the queuing policy can only monitor the system infrequently.  Infrequent monitoring will result
in a possibility of packets being served past their
deadline. Thus, to prevent that, dropping policies will
have to drop packets before their deadline. However,
by dropping packets prematurely, we can lose packets
that could have made their deadline. Analyzing this
tradeoff is the center of this paper.

More specifically, we will analyze a discrete-time M/M/1 queue where the policy can only drop packets after a packet arrival event happens. We assume all the packets have the same hard deadline ($D$). We use two different approaches to this problem.

As a first approach, we write the optimization problem as a Markov Decision Process (MDP). We find the optimal policies for small deadlines ($D=2,3$). We also mention some observations and properties of the optimal policies. 

Noting that finding the optimal policy for general deadline using the MDP approach is inefficient, we will provide a heuristic policy called $"AB-n"$. This policy reduces the computation complexity by only considering the first $n$ packets in the queue and maximizing their chances of meeting their respective deadlines. We will show that this policy outperforms the previously proposed algorithms.

 The rest of the paper is organized as follows. In section \ref{II}, we formulate the problem and discuss the notion of extended states and infrequent monitoring. In section \ref{III}, some previous results (namely EDF and DPGP) are mentioned. In section \ref{MDP}, we formulate the MDP for this problem. Section \ref{V} presents the results for the optimal policy stemming from the MDP. 
 The heuristic policy $AB-n$ is introduced in section \ref{Heuristic}.
 Section \ref{VII} presents some experiments. Section \ref{VIII} concludes the paper.  
\section{Problem Formulation}\label{II}
In this section, we formally define the problem. 
We will use the conventional discrete-time model for the M/M/1 queue described in \cite{mohanty1990}. The time is divided into time slots of unit duration, and we have the following rules:
\begin{itemize}
    \item At each time slot, at most one packet arrival or service happens. Arrival and service can not occur at the same time slot.
    \item Events in different time slots are independent.  
    \item At each time slot, an arrival happens with probability $\lambda$. If the queue is not empty, a packet service happens with probability $\mu$. 
\end{itemize}

Furthermore, all packets have the same hard deadline ($D$). Packets that do not meet the deadline are rendered pointless. We are trying to maximize the percentage of packets that meet their deadline. 

At each time, the queue is recorded as $T=(T_1,..., T_N)$ where $T_i$ is the age of the $i^{th}$ packet (i.e., the time elapsed since its arrival). We assume $T_1$ is the head of the queue and $T_N$ is the tail. This is the same definition presented at \cite{Farahvash2023}  as extensive states.

Finally, unlike previous results (i.e., EDF), the queue is only monitored after a packet arrival (infrequent monitoring). In other words, the policy is only allowed to drop a packet (or packets) after a packet arrives at the queue. Packets are only dropped from the head of the queue.

\section{Previous Results}\label{III}
In this section, we will present some previous results that will serve as points of comparison. More specifically, we will look into two specific policies (DPGP and EDF) and explain why these policies are not optimal in our model.

\subsection{Earliest Deadline First (EDF)}
In the real-time queuing literature, \cite{Panwar1988}
provides the following definition for the Earliest Deadline First (or Shortest Time to Extinction (STE)) Policies and proves that EDF is the optimal policy for the discrete-time G/M/c+G queue.
\begin{definition}
    A policy is an \textbf{Earliest Deadline First} policy if (1) at any time, it always schedules the available packet closest to its deadline, (2) the servers are always busy as long as there are available packets (there are no forced idle times), and (3) the packets are discarded (removed from the queue and server) as soon as their deadline elapses.
\end{definition}

\begin{remark}
    Note that the third condition for EDF policies can not be enforced with infrequent monitoring (the policy is only allowed to drop at certain time slots).  
\end{remark}
In the next part, we will provide an intuitive example to illustrate the reason why EDF can be sub-optimal with infrequent monitoring. 

\subsection{An intuitive Example}\label{Example}
Let's assume we are looking into the queue at time slot t. Also, let the packet at the head of the queue ($p_1$) and the second packet ($p_2$) have $k$ and $k+k'$ timeslots till their deadline, respectively. We are interested in the expected number of the packets making their deadline out of these two packets (we ignore the rest of the queue). We want to see whether dropping the packet at the head of the queue at time t can increase this probability.

First, we look into the constant monitoring case. By dropping the packet at the head of the queue at time $t$, the second packet will have $k+k'$ timeslots to be served. Thus, the expected number of packets served before their deadline is equal to $A=1-(1-\mu)^{k+k'}$. Now, by not dropping the packet, $p_1$ will have $k$ timeslots to be served. Furthermore, $p_2$ will have at least $k'$ timeslots to be served. (because of the constant monitoring $p_1$ is dropped after $k$ timeslots if not served.) Hence, the expected number of packets being served is at least $B=2-(1-\mu)^{k}-(1-\mu)^{k'}$. It is easy to see that $\forall k,k' \in \mathbb{N},\mu \in (0,1)$,  $B\geq A$. Thus, dropping a packet before its deadline never increases the expectation (regardless of the values of $t,k,k',\mu$).

Now, we go back into the infrequent monitoring case. Here, by dropping the packet at the head of the queue at time $t$, the desired expected value will be the same as the constant monitoring case (i.e., $A$). Finally, if we decide to not drop the packet at time $t$, $p_1$ will still have $k$ timeslots to be served. But contrary to the frequent monitoring case, $p_2$ can have less than $k'$ timeslots to be served (depending on the next arrival and departure times). Thus, there exist cases where dropping will increase the expectation. We will inspect these cases more thoroughly in future sections.
\subsection{Drop Positive Gain Policy (DPGP)}
  \cite{Farahvash2023} argues that when deciding to drop a packet, two factors should be considered:
\begin{enumerate}
    \item The probability of the packet making the deadline
    \item The impact of the packet being dropped on the probability of other packets meeting their respective deadlines.
\end{enumerate}
\begin{remark}\label{Incomplete Beta}
    The probability of the $i^{th}$ packet making the deadline is equal to $I_\mu(i, D-T_i-i+1)$, Where $I$ is the regularized incomplete beta function \cite{Abramowitz1988} (Section 6.6).
\end{remark}
To formally compute the trade-off of the two factors, the paper introduces the concept of gain as follows (The original definition is for the continuous time queue. We change the definition to fit the discrete model).

\begin{definition}
    The gain of dropping the $i^{th}$ packet in state s is defined as:
 \begin{align}
        & gain^s_i(\mu) =-I_\mu(i,D-T_i-i+1)\\
        &+\sum_{j=i+1}^{N} \left(I_\mu(j-1,D-T_j-j+2)-I_\mu(j,D-T_j-j+1)\right)\nonumber   
\end{align}
\end{definition}
Using the gain function defined above and by proving that the gain function is maximized at the head of the queue, \cite{Farahvash2023} introduces the Drop Positive Gain Policy as follows:

"Drop the packet at the head of the queue if and only if the $gain_1^s(\mu) >0 $."

\begin{remark}
    $gain$ is a myopic concept and ignores future system evolution (packet arrival, service, or drop). Thus, DPGP is not the optimal policy.
\end{remark}

Now that we have established that the previously proposed policies are not optimal, we will provide two different approaches to finding better policies. First, in section \ref{MDP}, we will try to find the optimal policy by formulating the queue as an MDP. Next, in section \ref{Heuristic}, using the intuition gained from the example above, we propose a heuristic policy.

\section{Queue as an MDP}\label{MDP} 
Markov Decision Processes (MDPs) are used for
making a sequence of decisions in situations where
the outcomes are uncertain. This framework can fit the
queue optimization problem that we are analyzing in
this paper. The MDP we use for this problem is an
infinite-horizon MDP where we are trying to maximize
the average reward per stage
\subsection{General definition}
An MDP is defined as $\mathcal{M}=(S,\mathcal{A},P,r)$ where:
\begin{itemize}
    \item $S$ is the state space.
    \item $\mathcal{A}$ is the action space. It maps a state to admissible actions for that state.\\ In other words $\mathcal{A}: S \rightarrow \{actions\} $.
    \item P is the transition function. $P:S \times \mathcal{A} \rightarrow \Delta(S)$ where $\Delta(S)$ is the space of probability distributions over S. In simple words, P explains the transition probability from a state and an action taken when in that state, to a new state.
    \item r is the reward function. $r:S \times \mathcal{A} \rightarrow \mathbf{R}$. $r(s,a)$ is the stage reward associated with taking action $a$ in state $s$. In a more general framework, $r(s,a)$ can be a random variable.
\end{itemize}
We will try to optimize the average reward per stage starting from state s (using policy $\pi$), which is defined as:
\begin{equation*}
    J^{\pi}(s) = \lim_{N \rightarrow \infty} \frac{1}{N} \mathbf{E}\left [\sum_{t=0}^{N-1} r(s_t,a_t) | \pi,s_0=s \right]
\end{equation*}
In the next part, we will formulate the problem of maximizing the expected percentage of packets meeting the deadline as an MDP.
\subsection{MDP Design}\label{}
To derive the MDP for this problem, we will define each of $(S,\mathcal{A}, P,r)$. We also note that state transitions will happen after a packet arrives, is served, or is dropped. 
\subsubsection{S} The states are the same as $T$ (extensive states) with two additional binary bits $(b_a,b_r)$ which are defined as below:
\begin{equation}
    b_a=\begin{cases}
        1, & \text{If we are allowed to drop packets}\\
        0, & \text{otherwise}
    \end{cases}
\end{equation}
And:
\begin{equation}
    b_r=\begin{cases}
        1, & \text{If $b_a=0$ and the previously served } \\
        &\text{packet made the deadline}\\
        0, & \text{otherwise}
    \end{cases}
\end{equation}
In other words, $b_a$ accounts for whether or not a packet service was the reason for the previous state transition (this will help us define the action space). $b_r$ shows that if the previous state transition was caused by a packet service, whether the packet made the deadline or not (this will help us define the reward function).

To conclude, state s can be defined as: 
\begin{equation*}
    s=((T^s_1,...,T^s_n),b^s_a,b^s_r)
\end{equation*}
\subsubsection{$\mathcal{A}$} To define the action space, we will use $b_a$. We have two possibilities:
\begin{itemize}
    \item $b^s_a=0$: If $b^s_a=0$, a packet service was the reason for the previous state transition. Thus, we are not allowed to drop a packet and:
    \begin{equation*}
        \mathcal{A}(s)= \{\bar{d} \}
    \end{equation*}
    Where $\Bar{d}$ means not dropping a packet.
    \item $b^s_a=1$: If $b^s_a=1$, a packet service was not the reason for the previous state transition. Thus, we are allowed to drop a packet and:
    \begin{equation*}
        \mathcal{A}(s)= \{\bar{d}, d \}
    \end{equation*}
    Where $\Bar{d}$ means not dropping a packet, and $d$ means dropping the packet at the head.
\end{itemize}
\subsubsection{P} We will first provide the following lemma. \begin{lemma}\label{lemma 3}
    If the queue is not empty, the interval between two consecutive state transitions comes from a geometric distribution with parameter $\lambda+\mu$ and the probability of an arrival triggering the state transition is $\frac{\lambda}{\lambda+\mu}$.       
\end{lemma}
\begin{proof}
    If the queue is not empty, the probability of no arrival and no service (failure) in a slot is $1-\mu-\lambda$. Thus, the probability distribution of the first success (arrival or packet service) is $geom(\lambda+\mu)$. The probability of the success being from arrival is $\frac{\lambda}{\lambda+\mu}$.  
\end{proof}
\begin{remark}
     If the queue is empty, the interval between two consecutive state transitions comes from a geometric distribution with parameter $\lambda$.
\end{remark}
To describe the transition function, we condition it on the action taken:
\begin{itemize}
    \item $a=d$: The next state will be:
    \begin{equation*}
        s'=((T^s_2,\ldots,T^s_n),1,0)
    \end{equation*}
    with probability 1. \item $a=\Bar{d}$: Here the next state is stochastic. We have:
    \begin{itemize}
        \item n=0: Then the next state will be 
        \begin{equation*}
             s'=((0),1,0)
        \end{equation*}
        with probability 1.
        \item $n>0$: Then using lemma \ref{lemma 3}, we get that with probability $\frac{\lambda}{\lambda+\mu}$:
        \begin{equation*}
             s'=((T^s_1+K,...,T^s_n+K,0),1,0)
        \end{equation*}
        where $K \sim geom(\lambda+\mu)$.
        
        Otherwise, with probability $\frac{\mu}{\lambda+\mu}$:
        \begin{equation*}\label{service}
             s'=((T^s_2+K,...,T^s_n+K),0,b^{s'}_r)
        \end{equation*}
        where $K \sim geom(\lambda+\mu)$ and:
        \begin{equation*}
            b^{s'}_r=
            \begin{cases}
                1, & \text{If $K \leq D-T^s_1$}\\
                0, & \text{Otherwise}
            \end{cases}
        \end{equation*}
    \end{itemize}
\end{itemize}
\subsubsection{r} The reward function is pretty simple:
\begin{equation*}
    r(s,a)=2b^s_r
\end{equation*}
In the next part, we will provide reasoning on the equivalence of the MDP defined above and the optimization problem itself.
\subsection{On the equivalence of the MDP and original problem}
Note that we are optimizing:
\begin{equation*}
    J^{\pi}(s) = \lim_{N \rightarrow \infty} \frac{1}{N} \mathbf{E}\left [\sum_{t=0}^{N-1} r(s_t,a_t) | \pi,s_0=s \right]
\end{equation*}
Now, first, note that each packet causes two state transitions (once with arrival and once with departure). Thus, asymptotically speaking, if we let the number of packets till state transition number N be $Y_N$, we have:
\begin{equation*}
    \lim_{N \rightarrow \infty} \frac{Y_N}{N} = \frac{1}{2}
\end{equation*}
and thus, we can rewrite $J^{\pi}(s)$:
\begin{equation*}
    J^{\pi}(s) = \lim_{N \rightarrow \infty} \frac{1}{2Y_N} \mathbf{E}\left [\sum_{t=0}^{N-1} 2b^{s_t}_r | \pi,s_0=s \right]
\end{equation*}
$\sum_{t=0}^{N-1} b^{s_t}_r$ is equal to the number of packets meeting the deadline till N. Thus:
\begin{equation*}
    \lim_{N \rightarrow \infty} \frac{1}{Y_N} \mathbf{E}\left [\sum_{t=0}^{N-1} b^{s_t}_r | \pi,s_0=s \right] = \mathbf{P}(\text{meeting the deadline})
\end{equation*}
Thus, an optimal policy for the MDP problem maximizes the expected percentage of packets meeting the deadline. Furthermore, the reward incurred by a policy on the MDP is the same as the probability of a packet meeting the deadline if policy $\pi$ is implemented for the original problem. 

Thus, to the extent of our interest, these two problems are equivalent. 
\section{MDP optimization}\label{V}
In this section, we present results on the optimal policies for the MDPs presented in previous sections.

 As the number of states is infinite (or with some considerations mentioned in the next part, exponential), it is not efficient to solve this MDP using methods such as policy iteration or value iteration. But, we will provide the optimal policy for some special cases (i.e., $D=2,3$)

\subsection{The case D=2}
Here, we would define the optimal policy for the special case where the deadline is equal to 2. Any optimal policy would drop the packets that have missed the deadline (If they are allowed to drop packets). Furthermore, the queue under any optimal policy will never have a length of more than 3. Finally, if $T_i\geq D$, we will say $T_i=D$ since we only care that the packet has missed the deadline.  
Also, for simplicity define $\alpha=\lambda+\mu$.

Using the above considerations, the queue for D=2 will have 9 possible states. We will describe the state transitions here:
\begin{enumerate}
    \item $((),0,0)$: The only possible action is not dropping, and the next state will be $((0),1,0)$ with probability 1.
    \item $((),0,1)$: The only possible action is not dropping, and the next state will be $((0),1,0)$ with probability 1.
    \item $((0),1,0)$: The only possible action is not dropping (as if we drop the packet here, we will circulate between state 1 and this state forever, and the reward would be zero.). We have:
     \begin{equation*}
            ((0),1,0) \xrightarrow{\Bar{d}}
            \begin{cases}
                ((),0,0), & \text{wp $\mu \frac{(1-\alpha)^2}{\alpha}$}\\
                ((),0,1), & \text{wp $\mu (2-\alpha)$} \\
                ((1,0),1,0), & \text{wp $\lambda$}\\
                ((2,0),1,0), & \text{wp $\lambda \frac{(1-\alpha)}{\alpha}$} 
            \end{cases}
    \end{equation*}
    \item $((1),0,1)$: The only possible action is not dropping, and the next state will be:
    \begin{equation*}\label{next state}
            ((1),0,1) \xrightarrow{\Bar{d}}
            \begin{cases}
                ((),0,0), & \text{wp $\mu \frac{(1-\alpha)}{\alpha}$}\\
                ((),0,1), & \text{wp $\mu$} \\
                ((2,0),1,0), & \text{wp $ \frac{\lambda}{\alpha}$} 
            \end{cases}
    \end{equation*}
    \item $((2),0,0)$: The only possible action is not dropping, and the next state will be:
    \begin{equation*}
            ((2),0,0) \xrightarrow{\Bar{d}}
            \begin{cases}
                ((),0,0), & \text{wp $\frac{\mu}{\alpha}$}\\
                
                ((2,0),1,0), & \text{wp $ \frac{\lambda}{\alpha}$} 
            \end{cases}
    \end{equation*}
    \item $((1,0),1,0)$: This is the most important state, and we have two possible actions:
    \begin{itemize}
        \item Drop packet 1: The next state will be $((0),1,0)$ with probability 1.
        \item Don't drop: The next will be:
        \begin{equation*}
            ((1,0),1,0) \xrightarrow{\Bar{d}}
            \begin{cases}
                ((1),0,1), & \text{wp $\mu$}\\
                ((2),0,0), & \text{wp $\mu\frac{1-\alpha}{\alpha}$}\\
                ((2,1,0),1,0), &\text{wp $\lambda$}\\
                ((2,2,0),1,0), & \text{wp $ \frac{\lambda(1-\alpha)}{\alpha}$} 
            \end{cases}
    \end{equation*}
    \end{itemize}
    \item $((2,0),1,0)$: Any optimal policy would drop packet 1 as it has missed its deadline. Thus, the next state will be $((0),1,0)$.
    \item $((2,1,0),1,0)$: Any optimal policy would drop packet 1 as it has missed its deadline. Thus, the next state will be $((1,0),1,0)$.
    \item $((2,2,0),1,0)$: Any optimal policy would drop packet 1 as it has missed its deadline. Thus, the next state will be $((2,0),1,0)$.
\end{enumerate}
Thus, any policy has to decide which action to take in state 6. Let's see what decision DPGP makes. $gain^s_1(\mu)$ is equal to:
    \begin{equation*}
        gain^s_1(\mu)= (\mu+\mu(1-\mu))-\mu^2-\mu = \mu - 2 \mu^2
    \end{equation*}
    Thus, DPGP will be:
    \begin{equation}
        a(s_6)=\begin{cases}
                d, & \text{if $\mu < 0.5$}\\
                \Bar{d}, & \text{if $\mu \geq 0.5$}
            \end{cases}
    \end{equation}

To compute the optimal policy, let $\pi^d$ be the stationary distribution of the Markov chain induced by policy $d$ on our MDP. Now, by definition of the reward ($r=2b_r$), the percentage of packets meeting the deadline will be $2(\pi_2^d+\pi_4^d)$.

Thus, given $\lambda,\mu$, the optimal policy maximizes 
\\$2(\pi_2^d+\pi_4^d)$ (Call that $AR(\lambda,\mu)$).

Depending on $\mu$ and $\lambda$, the optimal policy has one of these two forms:
\begin{itemize}
    \item $a(s_6)=d$
    \item $a(s_6)=\Bar{d}$
\end{itemize}

\begin{figure}[b]
\centering
\includegraphics[width=0.8\linewidth]
{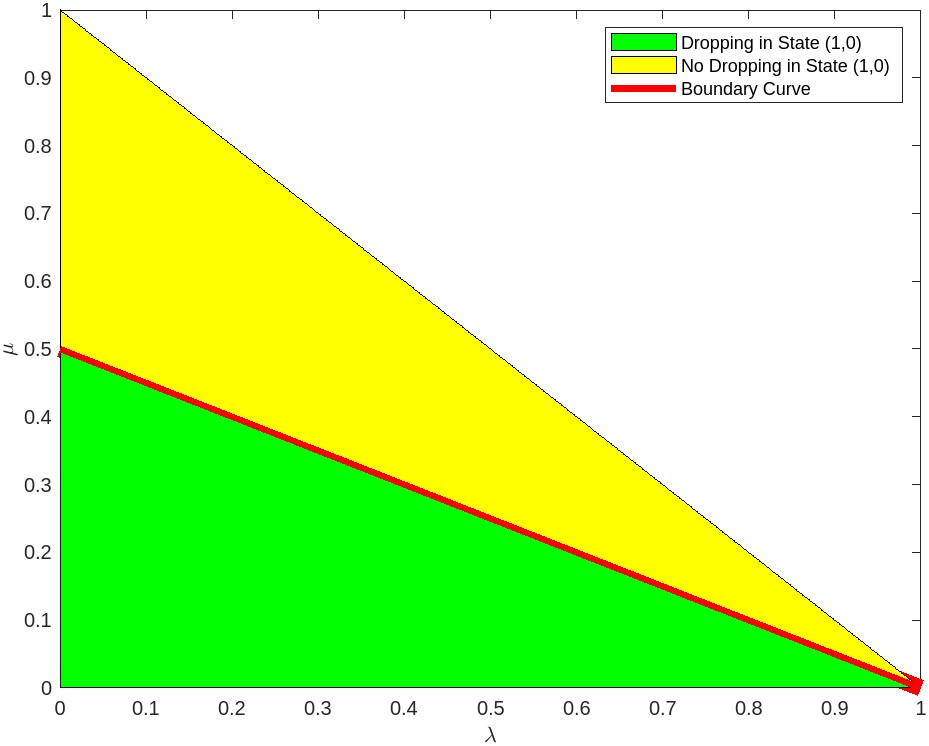}
\caption{Optimal policy for D=2}
\label{D2}
\end{figure}

We will compute the rewards of each policy and compute the optimal policy.
\begin{enumerate}
    \item $a(s_6)=d$: If we decide to drop from the head in state 6, the Markov chain will have the structure shown in Fig. \ref{fig:Markov}.
	
 \begin{figure}[b]
 \label{fig:Markov}
 \centering
  \begin{tikzpicture}[->, >=stealth', auto, semithick, node distance=3cm]
    
	\tikzstyle{every state}=[fill=white,draw=black,thick,text=black,scale=1]
	\node[state]    (A)                     {$1$};
	\node[state]    (B)[right of=A]   {$2$};
	\node[state]    (C)[below of=A]   {$6$};
	\node[state]    (D)[right of=C]   {$3$};
        \node[state]    (E)[right of=D]   {$7$};
	\path
	(A) edge[bend left,above]			node{$1$}(D)
	(B) edge[ right,left]	node{$1$}	(D)
	(C) edge[bend left,below]	node{$1$}	(D)
	(E) edge[bend right]		node{$1$}	(D)
        (D) edge[left,left]		node{$\frac{(1-\alpha)^2\mu}{\alpha}$}	(A)
        edge[bend right, right]		node{$\mu(2-\alpha)$}	(B)
        edge[bend left, below]		node{$\lambda$}	(C)
        edge[bend right, below]		node{$\lambda\frac{1-\alpha}{\alpha}$}	(E)
        ;
	\end{tikzpicture}
 \caption{The Markov chain of $a=d$}
 \end{figure}
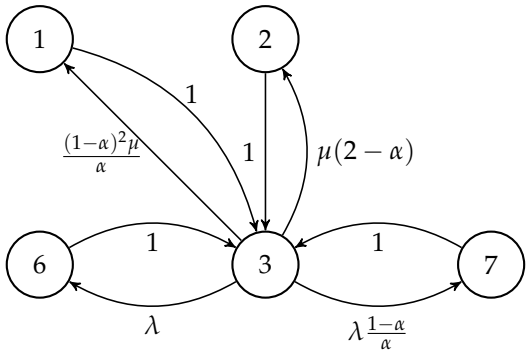
 The transition matrix would have the format (deleting the states that we will never enter ):
 \begin{equation*}
    P=
     \begin{bmatrix}
     1:&0&0&1&0&0 \\
     2:&0&0&1&0&0 \\
    3:&\mu\frac{(1-\alpha)^2}{\alpha} &\mu(2-\alpha)&0&\lambda& \lambda\frac{1-\alpha}{\alpha}\\
     6:&0&0&1&0&0 \\
     7:&0&0&1&0&0 \\
     \end{bmatrix}
 \end{equation*}
 To compute the stationary distribution, we must have $\pi P=\pi$. 
 We have:
 \begin{align}
     &\pi_1+\pi_2+\pi_3+\pi_6+\pi_7=1 \label{con1}\\
     &\pi_3=\pi_1+\pi_2+\pi_6+\pi_7 \label{con2}\\
     &\pi_2=\mu(2-\alpha)\pi_3 \label{con3}
 \end{align}
 Combining equation \ref{con1} and \ref{con2}, we get that $\pi_3=0.5$ and plugging it in equation \ref{con3}, we get:
 \begin{equation}
     AR_1(\lambda,\mu)=2(\pi_2^d+\pi_4^d)=\mu(2-\alpha)
 \end{equation}
 \item $a(s_6)=\Bar{d}$: If we decide not to drop from the head in state 6, the Markov chain will have the following transition matrix
 \footnote{
 \begin{equation*}
     P=\begin{bmatrix}
     0&0&1&0&0&0&0&0&0 \\
     0&0&1&0&0&0&0&0&0 \\
    \mu\frac{(1-\alpha)^2}{\alpha} &\mu(2-\alpha)&0&0&0&\lambda& \lambda\frac{1-\alpha}{\alpha}&0&0\\
     \mu\frac{(1-\alpha)}{\alpha} &\mu&0&0&0&0& \frac{\lambda}{\alpha}&0&0\\
     \frac{\mu}{\alpha}&0&0&0&0&0&\frac{\lambda}{\alpha}&0&0 \\
     0&0&0&\mu&\mu\frac{(1-\alpha)}{\alpha}&0&0&\lambda&\lambda\frac{1-\alpha}{\alpha} \\
     0&0&1&0&0&0&0&0&0 \\
     0&0&0&0&0&1&0&0&0 \\
     0&0&0&0&0&0&1&0&0 \\
     \end{bmatrix}
 \end{equation*}}.
 Without going into further detail, we present the stationary distribution for important states:
 \begin{align*}
     &\pi_2=\frac{\mu}{2}[(2-\alpha)(1-\lambda)+\mu],
     &\pi_4=\frac{\mu\lambda}{2}
 \end{align*}
 We have:
 \begin{equation}
     \begin{aligned}
          AR_2(\lambda,\mu) &= 2(\pi_2^d+\pi_4^d)\\ 
     &=\mu(2-\alpha) + \lambda\mu[2\mu+\lambda-1]
     \end{aligned}
 \end{equation}
\end{enumerate}
Thus, the optimal policy would decide to drop the packet in state 6 if and only if $AR_1(\lambda,\mu)>AR_2(\lambda,\mu)$. This will happen when:
\begin{equation*}
    AR_1(\lambda,\mu)>AR_2(\lambda,\mu) \leftrightarrow 2\mu+\lambda<1
\end{equation*}
Therefore, the optimal policy is described below:
\begin{equation}
        a^*(s_6)=\begin{cases}
                d, & \text{if $2\mu+\lambda < 1$}\\
                \Bar{d}, & \text{Otherwise}
            \end{cases}
\end{equation}
Fig. \ref{D2} shows the boundary of the optimal policy. For any pair ($\lambda,\mu$) below the red line, the optimal policy would drop from the head when at state (1,0). Similarly,
for any pair ($\lambda,\mu$) above the red line, the optimal policy wouldn't drop any packets at state (1,0).

\begin{remark}
    Note that DPGP is the same as the optimal policy for $\lambda=0$. This is true as DPGP ignores any effect that a new arrival has on the system. Thus, for small arrival rates, the myopic gain computed by DPGP is close to the actual gain.
\end{remark}
\subsection{The case D=3}\label{SectionD3}
Here, we will define the optimal policy for the special case where the deadline is equal to 3. We have the same considerations as D=2. Without going into further detail (for a complete analysis, visit Appendix A), we mention that the system will have 20 states, and the optimal policy should decide in 3 different states (that are non-trivial) whether to drop from the head or not drop at all.
\begin{itemize}
    \item ((1,0),1,0) 
    \item ((2,0),1,0)
    \item ((2,1,0),1,0)
\end{itemize}
We see that, depending on $(\lambda,\mu)$, the optimal policy has one of the following forms:
\begin{enumerate}[(a)]
    \item Drops in all of the above states.
    \item Drops in states ((2,1,0),1,0) and ((2,0),1,0).
     \item Only drops in state ((2,1,0),1,0).
    \item Drops in none of the above states.
\end{enumerate}

\begin{figure}[b]
\centering
\includegraphics[width=0.9\linewidth]
{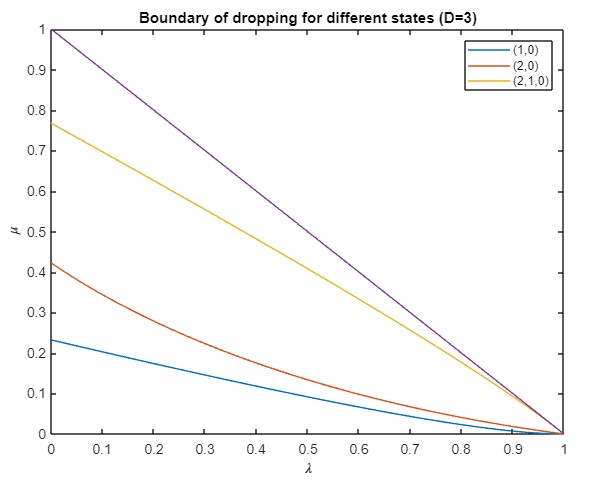}
\caption{optimal policy boundary for D=3}
\label{D3}
\end{figure}

Fig. \ref{D3} shows the boundaries of the optimal policy. (For the computation of these boundaries, visit Appendix A). For any pair $(\lambda, \mu)$ below the blue line, the optimal policy would be (a). For $(\lambda, \mu)$ between red and blue lines, policy (b) would be optimal. If the point is between red and yellow lines, we would only drop in the state $((2,1,0),1,0)$ (i.e., policy (c)). Finally, if we are above the yellow line, (d) is the optimal policy.

Alternatively, for any point below the yellow line, we would drop at state ((2,1,0),1,0). If $(\lambda,\mu)$ is below the red line, an optimal policy would drop at state ((2,0),1,0), and for all parameters below the blue line, we would drop at state ((1,0),1,0). 

\subsection{On properties of the boundaries}
In this section, we will present some key properties of the boundaries of the optimal policies for general deadline D. (The boundaries for special cases can be seen in Fig. \ref{D2}, \ref{D3}, and \ref{D4})

From now on, for any state $s$, we will refer to the boundary point at $\lambda=0$ as $M(s)$. Also, we let $\mu_{bound}$ to be the service rate where $gain_1^s(\mu^s_{bound})=0$.
\begin{theorem}\label{myopic theorem}
    For any state s, $M(s)=\mu^s_{bound}$.
\end{theorem}
\begin{proof}
    If the arrival rate is equal to zero, we have two important observations:
    \begin{enumerate}
        \item The reward only depends on the packets already in the system (there will not be packet arrivals).
        \item As there are no arrivals in the future, we can not drop any packets in the future, and thus the probability of the $i^{th}$ packet making the deadline is exactly $I_\mu(i,D-T_i-i+1)$ if we decide to keep the packet at the head and $I_\mu(i-1,D-T_i-i+2)$ if we decide to drop. (\ref{Incomplete Beta}) 
    \end{enumerate}
    Thus, the M(s) would be the service rate where the gain is indifferent to the dropping policy at $s$. In other words:
    \begin{equation*}
        \sum_{j=1}^{N} I_{M(s)}(j,D-T_j-j+1) = \sum_{j=2}^{N} I_{M(s)}(j-1,D-T_j-j+2)
    \end{equation*}
    By a little rearrangement, we get:
    \begin{align*}
        &\sum_{j=2}^{N} \left(I_{M(s)}(j-1,D-T_j-j+2) - I_{M(s)}(j,D-T_j-j+1)\right) \nonumber\\& - I_{M(s)}(1,D-T_1) =0
    \end{align*}
    Which means:
    \begin{equation}
        gain_1^s(M(s))=0 \rightarrow M(s)=\mu^s_{bound}
    \end{equation}
\end{proof}
Finally, we will present a conjecture we suspect to be true regarding the boundary lines.
\begin{conjecture}\label{main}
 For a given deadline D, non-trivial states have an ordering. More specifically, let $B^{s}(\lambda)$ be the boundary curve for the optimal policy at non-trivial state $s$. These curves don't cross. More precisely:
    \begin{equation}
        \nexists s_1,s_2 \in S, \lambda<1: B^{s_1}(\lambda)=B^{s_2}(\lambda) , B^{s_1}(\lambda)\neq 1-\lambda
    \end{equation}
\end{conjecture}

\section{Heuristic Policy ($AB-n$)}\label{Heuristic}
As established in the previous section, finding the optimal policy using the MDP method is not feasible for larger deadlines. Thus, in this section, we will introduce a heuristic algorithm to approximate the optimal policy. 

To do that, we will look back into the example provided in section \ref{Example}. Recall that we are looking into the first two packets at the head of the queue (with $k$ and $k+k'$ timeslots until expiration), and we are trying to maximize the expected number of packets meeting their deadline out of the two packets. 

If we decide to drop the packet at the head of the queue, the expectation is equal to $A_2(k,k')=1-(1-\mu)^{k+k'}$. Now, we compute the expectation without dropping the packet. Call this number $B_2(k,k')$. We calculate this value by conditioning on the next event (arrival or departure). There are three types of possibilities for the next event.
\begin{enumerate}
    \item An arrival event happens before the deadline of the first packet: In this case, the expected value is equal to $max(A_2(k-i,k'),B_2(k-i,k'))$ where $i$ is the time of the packet arrival.
    \item A departure event happens before the deadline of the first packet: In this case, the expected value is equal to $2-(1-\mu)^{k+k'-i}$ where $i$ is the time of the packet departure.
    \item An arrival or departure event happens after the deadline of the first packet: In this case, the expected value is equal to $1-(1-\mu)^{k+k'-i}$ where $i$ is the time of the event happening.
\end{enumerate}
Putting all the results above together, we get:
\begin{equation}
    \begin{aligned}
        &B_2(k,k')=\\
        &\sum_{i=1}^k \lambda (1-\lambda-\mu)^{i-1} \underbrace{\max(A_2(k-i,k'),B_2(k-i,k'))}_1 \\
        &+\sum_{i=1}^k \mu (1-\lambda-\mu)^{i-1} \underbrace{(2-(1-\mu)^{k+k'-i})}_2\\
        &+\sum_{i=k+1}^{k+k'} (\mu+\lambda) (1-\lambda-\mu)^{i-1} \underbrace{(1-(1-\mu)^{k+k'-i})}_3\\
    \end{aligned}
\end{equation}
\begin{remark}
   Note that $B_2(k,k')$ only depends on $B_2(l,k')$ with $l \leq k$. Thus, we can calculate $B_2(k,k')$ without the need to solve linear equations.     
\end{remark}
Now, that we have calculated $B_2(k,k')$, we can introduce the "$AB-2$" policy:\\
\textbf{$\mathbf{AB-2}$ Policy:} While deciding whether to drop in state T, the $AB-2$ policy will drop the packet if and only if either $T_1\geq D$ or $|T|>1$ and $A_2(D-T_1,T_2-T_1)>B_2(D-T_1,T_2-T_1)$.

\begin{remark}
       If the deadline is equal to 2 (D=2), the $AB-2$ policy would drop in state $T=(1,0)$ (the only nontrivial state) iff $\lambda+2\mu<1$ which is the same as the optimal policy.
\end{remark}
Note that by extending the number of packets considered in the expected value, we can improve the $AB-2$ policy (We call it the $Ab-n$ policy where n is the number of packets considered). $A_3$ and $B_3$ for $AB-3$ policy can be seen in equations \ref{A-3} and \ref{B-3} respectively. 
\begin{table*}
\centering
\begin{minipage}{0.9\textwidth}
\begin{equation}\label{A-3}
A_3(k_1,k_2,k_3)=\max(A_2(k_1+k_2,k_3),B_2(k_1+k_2,k_3))
\end{equation}
\begin{equation}\label{B-3}
    \begin{aligned}
        B_3(k_1,k_2,k_3)&=\sum_{i=1}^{k_1} \lambda (1-\lambda-\mu)^{i-1} \max{(A_3(k_1-i,k_2,k_3),B_3(k_1-i,k_2,k_3)})
        +\sum_{i=1}^{k_1} \mu (1-\lambda-\mu)^{i-1} (1+B_2(k_1+k_2-i,k_3))\\
        &+\sum_{i=k_1+1}^{k_1+k_2} \lambda (1-\lambda-\mu)^{i-1} \max{(B_2(k_1+k_2-i,k_3),A_2(k_1+k_2-i,k_3)})+\sum_{i=k_1+1}^{k_1+k_2} \mu (1-\lambda-\mu)^{i-1} B_2(k_1+k_2-i,k_3)\\
         &+\sum_{i=k_1+k_2+1}^{k_1+k_2+k_3} \lambda (1-\lambda-\mu)^{i-1} (1-(1-\mu)^{k_1+k_2+k_3-i})+\sum_{i=k_1+k_2+1}^{k_1+k_2+k_3}  \sum_{j=i+1}^{k_1+k_2+k_3}(\mu^2+\mu\lambda)(1-\lambda-\mu)^{j-2}(1-(1-\mu)^{k_1+k_2+k_3-j})\\
    \end{aligned}
\end{equation} 
\hrule
\end{minipage}
\end{table*}

\section{Numerical results and experiments}\label{VII}
In this section, we will present some numerical examples and experiments pertaining to the results of this paper.
\subsection{The D=4 case}
\begin{figure}[b]
\centering
\includegraphics[width=0.8\linewidth]
{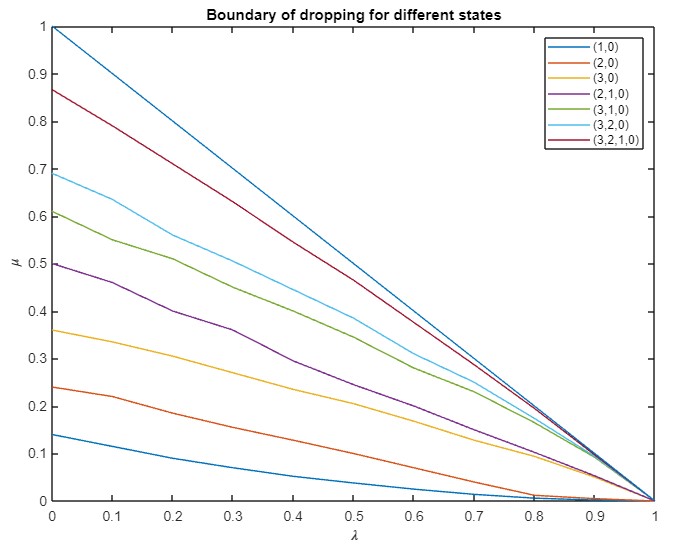}
\caption{Optimal policy boundary for D=4}
\label{D4}
\end{figure}

Here, we will experimentally find the optimal policy for the case where the deadline is equal to 4. In this case, we have 7 non-trivial states where our policy has to decide. To find the optimal policy, we implement the M/M/1 queue with different $\lambda$ and $\mu$ using each policy and find the policy that maximizes the percentage of packets meeting their deadline. Fig. \ref{D4} shows the boundary of the optimal policies. For any pair ($\lambda,\mu$) below each line, the optimal policy would drop from the head when at that state. For instance, if ($\lambda,\mu$) is below the green line, we would drop at state $((3,1,0),1,0)$.

\subsection{$AB-n$ with different n values}
\begin{figure}
\centering
\includegraphics[width=0.9\linewidth]
{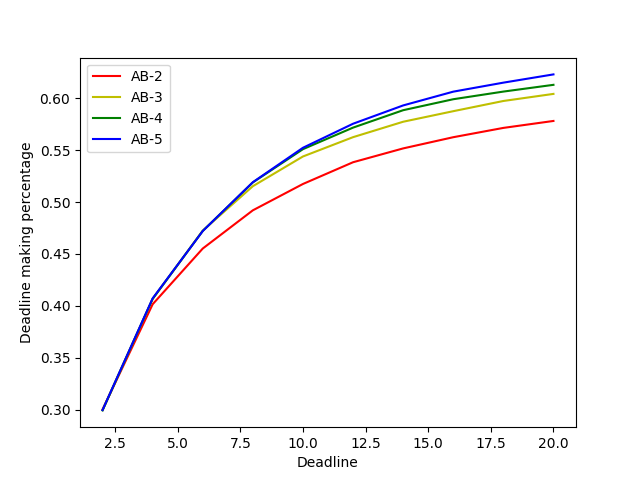}
\caption{$AB-n$ Performance}
\label{AB-n}
\end{figure}

Fig. \ref{AB-n}, compares the performance of $AB-n$ policies with different values of $n$ for $\lambda=0.3,\mu=0.2$. As can be seen, the percentage of packages meeting their deadlines improves by increasing the value of $n$, but the rate of improvement decreases as $n$ gets larger.

\subsection{$AB-5$ vs DPGP and EDF}
\begin{figure}
\centering
\includegraphics[width=0.9\linewidth]
{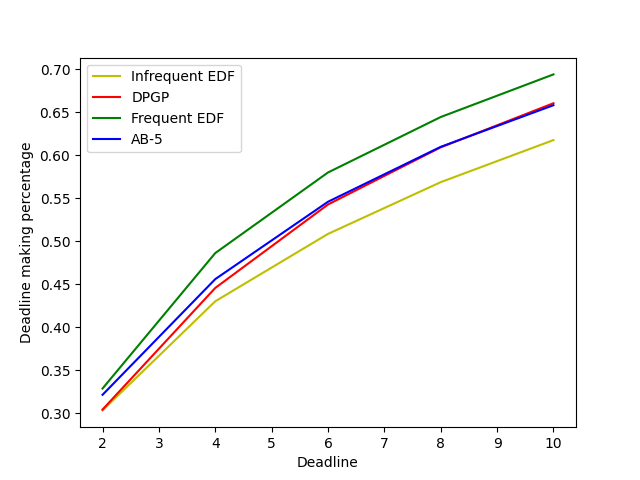}
\caption{$AB-n$ performance compared with previous algorithms}
\label{EDF}
\end{figure}

Fig. \ref{EDF} compares the performance of $AB-5$ policy with DPGP and EDF introduced in section \ref{III}. Both EDF with frequent and infrequent monitoring are used. $AB-5$ outperforms both DPGP and infrequent EDF for shorter deadlines. For larger deadlines, the performance of DPGP and $AB-5$ becomes almost identical. Generally speaking, we observe that, by increasing the deadline, DPGP will eventually outperform $AB-n$ (albeit slightly). But, by increasing $n$ this phenomenon happens later.
\section{Conclusion}\label{VIII}
In this paper, we looked into discrete time M/M/1 queues where packets have a hard deadline. We assumed that continuous monitoring of the system is not feasible. Thus, we introduced infrequent monitoring, where the system is only monitored after a packet arrival event happens. We tried to maximize the percentage of packets meeting their deadline.

We had two approaches to this problem. First, the queue was modeled as an MDP. We presented the optimal policy for small deadlines (D=2,3). Some properties of the optimal policies were discussed. 

As a second approach, we introduced a heuristic policy ($AB-n$) which improves the performance of the queue compared with previous algorithms (DPGP and EDF). Finally, some numerical simulations were provided to verify the results.

Also, it is worth noting that, as a line of future works,
this approach can be applied to latency based utility
optimization (cases without a hard deadline).
\bibliographystyle{ieeetr}
\bibliography{references}
\clearpage
\section*{Appendix A: The D=3 case}
Here, we provide a complete analysis of the $D=3$ case. Let's remember that the optimal policy should decide in 3 different states (that are nontrivial) whether to drop from the head or not drop at all.
\begin{itemize}
    \item ((1,0),1,0) 
    \item ((2,0),1,0)
    \item ((2,1,0),1,0)
\end{itemize}
\subsection{DPGP}
Once again, we will see what DPGP does in the states above:
\begin{itemize}
    \item s=((1,0),1,0): We have:
    \begin{equation*}
        gain^s_1(\mu)= \mu (1-5\mu + 3\mu^2)
    \end{equation*}
    We have:
    \begin{equation*}
       gain^s_1(\mu)>0 \leftrightarrow \mu < \frac{5-\sqrt{13}}{6} 
    \end{equation*}
    And:
    \begin{equation}
        a(s)=\begin{cases}
                d, & \text{if $\mu <\frac{5-\sqrt{13}}{6}  $}\\
                \Bar{d}, & \text{Otherwise}
            \end{cases}
\end{equation}
    \item s=((2,0),1,0):
    We have:
    \begin{equation*}
        gain^s_1(\mu)= \mu (2-6\mu + 3\mu^2)
    \end{equation*}
    We have:
    \begin{equation*}
       gain^s_1(\mu)>0 \leftrightarrow \mu < 1-\frac{\sqrt{3}}{3} 
    \end{equation*}
    And:
    \begin{equation}
        a(s)=\begin{cases}
                d, & \text{if $\mu <1-\frac{\sqrt{3}}{3}   $}\\
                \Bar{d}, & \text{Otherwise}
            \end{cases}
\end{equation}
    \item s=((2,1,0),1,0): We have:
    \begin{equation*}
        gain^s_1(\mu)= \mu (1+\mu - 3\mu^2)
    \end{equation*}
    We have:
    \begin{equation*}
       gain^s_1(\mu)>0 \leftrightarrow \mu < \frac{\sqrt{13}+1}{6} 
    \end{equation*}
    And:
    \begin{equation}
        a(s)=\begin{cases}
                d, & \text{if $\mu <\frac{\sqrt{13}+1}{6}  $}\\
                \Bar{d}, & \text{Otherwise}
            \end{cases}
\end{equation}
\end{itemize}
\subsection{States}
Here, we mention all 20 states for $D=3$:
\begin{enumerate}
    \item $((),0,0)$: The only possible action is not dropping, and the next state will be $((0),1,0)$ with probability 1.
    \item $((),0,1)$: The only possible action is not dropping, and the next state will be $((0),1,0)$ with probability 1.
    \item $((0),0,0)$: The only possible action is not dropping (as if we drop the packet here, we will circulate between state 1 and this state forever, and the reward would be zero.). We have:
     \begin{equation*}
            ((0),0,0) \xrightarrow{\Bar{d}}
            \begin{cases}
                ((),0,0), & \text{wp $\mu \frac{(1-\alpha)^3}{\alpha}$}\\
                ((),0,1), & \text{wp $\mu (1-(1-\alpha)^3)$} \\
                ((1,0),1,0), & \text{wp $\lambda$}\\
                ((2,0),1,0), & \text{wp $\lambda(1-\alpha)$}\\
                ((3,0),1,0), & \text{wp $\lambda \frac{(1-\alpha)^2}{\alpha}$} 
            \end{cases}
    \end{equation*}
    \item $((1),0,1)$: The only possible action is not dropping, and the next state will be:
    \begin{equation*}
            ((1),0,1) \xrightarrow{\Bar{d}}
            \begin{cases}
                ((),0,0), & \text{wp $\mu \frac{(1-\alpha)^2}{\alpha}$}\\
                ((),0,1), & \text{wp $\mu (2+\alpha)$} \\
                ((2,0),1,0), & \text{wp $\lambda$}\\
                ((3,0),1,0), & \text{wp $ \lambda\frac{1-\alpha}{\alpha}$} 
            \end{cases}
    \end{equation*}
    \item $((2),0,0)$: The only possible action is not dropping, and the next state will be:
    \begin{equation*}
            ((2),0,0) \xrightarrow{\Bar{d}}
            \begin{cases}
                ((),0,0), & \text{wp $\mu\frac{1-\alpha}{\alpha}$}\\
                ((),0,1), & \text{wp $\mu$}\\
                ((2,0),1,0), & \text{wp $ \frac{\lambda}{\alpha}$} 
            \end{cases}
    \end{equation*}
    \item $((2),0,1)$: The transition probability is the same as previous state.
    \item $((3),0,1)$: Again, we can not drop a packet here and:
    \begin{equation*}
            ((3),0,0) \xrightarrow{\Bar{d}}
            \begin{cases}
                ((),0,0), & \text{wp $\frac{\mu}{\alpha}$}\\
                ((3,0),1,0), & \text{wp $ \frac{\lambda}{\alpha}$} 
            \end{cases}
    \end{equation*}
    \item $((1,0),1,0)$: Similar to D=2, we have two possible actions:
    \begin{itemize}
        \item Drop packet 1: The next state will be $((0),1,0)$ with probability 1.
        \item Don't drop: The next will be:
        \begin{equation*}
            ((1,0),1,0) \xrightarrow{\Bar{d}}
            \begin{cases}
                ((1),0,1), & \text{wp $\mu$}\\
                ((2),0,1), & \text{wp $\mu(1-\alpha)$}\\
                ((3),0,0), & \text{wp $\mu\frac{(1-\alpha)^2}{\alpha}$}\\
                ((2,1,0),1,0), &\text{wp $\lambda$}\\
                ((3,2,0),1,0), & \text{wp $ \lambda(1-\alpha)$}\\
                ((3,3,0),1,0), & \text{wp $\lambda\frac{(1-\alpha)^2}{\alpha}$}
            \end{cases}
    \end{equation*}
    \end{itemize}
    \item $((2,0),1,0)$: Similar to the previous state, we have two possible actions:
    \begin{itemize}
        \item Drop packet 1: The next state will be $((0),1,0)$ with probability 1.
        \item Don't drop: The next will be:
        \begin{equation*}
            ((1,0),1,0) \xrightarrow{\Bar{d}}
            \begin{cases}
                ((1),0,1), & \text{wp $\mu$}\\
                ((2),0,0), & \text{wp $\mu(1-\alpha)$}\\
                ((3),0,0), & \text{wp $\mu\frac{(1-\alpha)^2}{\alpha}$}\\
                ((3,1,0),1,0), &\text{wp $\lambda$}\\
                ((3,2,0),1,0), & \text{wp $ \lambda(1-\alpha)$}\\
                ((3,3,0),1,0), & \text{wp $\lambda\frac{(1-\alpha)^2}{\alpha}$}
            \end{cases}
    \end{equation*}
    \end{itemize}
    \item $((2,1),0,1)$: The only possible action is not dropping, and the next state will be:
    \begin{equation*}
            ((2,1),0,1) \xrightarrow{\Bar{d}}
            \begin{cases}
                ((2),0,1), & \text{wp $\mu$}\\
                ((3),0,0), & \text{wp $\mu\frac{1-\alpha}{\alpha}$}\\
                ((3,2,0),1,0), & \text{wp $\lambda$}\\
                ((3,3,0),1,0), & \text{wp $ \lambda \frac{1-\alpha}{\alpha}$} 
            \end{cases}
    \end{equation*}
    \item $((3,0),1,0)$: Any optimal policy would drop packet 1 as it has missed its deadline. Thus, the next state will be $((0),1,0)$.
    \item $((3,2),0,0)$: The only possible action is not dropping, and the next state will be:
    \begin{equation*}
            ((3,2),0,0) \xrightarrow{\Bar{d}}
            \begin{cases}
                ((3),0,0), & \text{wp $\frac{\mu}{\alpha}$}\\
                ((3,3,0),1,0), & \text{wp $ \frac{\lambda}{\alpha}$} 
            \end{cases}
    \end{equation*}
     \item $((3,3),0,0)$: The only possible action is not dropping, and the next state will be:
    \begin{equation*}
            ((3,3),0,0) \xrightarrow{\Bar{d}}
            \begin{cases}
                ((3),0,0), & \text{wp $\frac{\mu}{\alpha}$}\\
                ((3,3,0),1,0), & \text{wp $ \frac{\lambda}{\alpha}$} 
            \end{cases}
    \end{equation*}
    \item $((2,1,0),1,0)$: we have two possible actions:
    \begin{itemize}
        \item Drop packet 1: The next state will be $((1,0),1,0)$ with probability 1.
        \item Don't drop: The next will be:
        \begin{equation*}
            ((2,1,0),1,0) \xrightarrow{\Bar{d}}
            \begin{cases}
                ((2,1),0,1), & \text{wp $\mu$}\\
                ((3,2),0,0), & \text{wp $\mu(1-\alpha)$}\\
                ((3,3),0,0), & \text{wp $\mu\frac{(1-\alpha)^2}{\alpha}$}\\
                ((3,2,1,0),1,0), &\text{wp $\lambda$}\\
                ((3,3,2,0),1,0), & \text{wp $ \lambda(1-\alpha)$}\\
                ((3,3,3,0),1,0), & \text{wp $\lambda\frac{(1-\alpha)^2}{\alpha}$}
            \end{cases}
    \end{equation*}
    \end{itemize}
     \item $((3,1,0),1,0)$: Any optimal policy would drop packet 1 as it has missed its deadline. Thus, the next state will be $((1,0),1,0)$.
      \item $((3,2,0),1,0)$: Any optimal policy would drop packet 1 as it has missed its deadline. Thus, the next state will be $((2,0),1,0)$.
       \item $((3,3,0),1,0)$: Any optimal policy would drop packet 1 as it has missed its deadline. Thus, the next state will be $((3,0),1,0)$.
        \item $((3,2,1,0),1,0)$: Any optimal policy would drop packet 1 as it has missed its deadline. Thus, the next state will be $((2,1,0),1,0)$.
         \item $((3,3,2,0),1,0)$: Any optimal policy would drop packet 1 as it has missed its deadline. Thus, the next state will be $((3,2,0),1,0)$.
          \item $((3,3,3,0),1,0)$: Any optimal policy would drop packet 1 as it has missed its deadline. Thus, the next state will be $((3,3,0),1,0)$.
    
\end{enumerate}
\subsection{Reward Computation} 
To compute the real optimal policy, let $\pi^d$ be the stationary distribution of the Markov chain induced by policy $d$ on our MDP. Now, by definition of the reward ($r=2b_r$), the percentage of packets meeting the deadline will be $2(\pi_2^d+\pi_4^d+\pi_6^d+\pi_{10}^d)$.

Thus, given $\lambda,\mu$, the optimal policy maximizes 
\\$AR(\lambda,\mu)=2(\pi_2^d+\pi_4^d+\pi_6^d+\pi_{10}^d)$. 

Similar to D=2, We will state the rewards of each policy (we will skip the computation of stationary distribution and mention the probability of the important states here).
Depending on $(\lambda,\mu)$, the optimal policy has one of the following forms:
\begin{enumerate}[(a)]
    \item Drops in all of the above states: If we decide to drop in all of the 3 nontrivial states, the important states in the stationary distribution will probabilities:
    \begin{equation*}
        \pi_2^d=\frac{\mu}{2\alpha}(1-(1-\alpha)^3), \pi_4^d=0, \pi_6^d=0, \pi_{10}^d=0
    \end{equation*}
    Thus, the reward of this policy would be:
    \begin{equation}
        AR_a(\lambda,\mu)=\mu(3-3\alpha+\alpha^2)
    \end{equation}
    \item Drops in states ((2,1,0),1,0) and ((2,0),1,0): If the policy decides to keep the packets at state (1,0) and drop them in states (2,0) and (2,1,0), we will have:
    \begin{equation*}
        \pi_2^d=\frac{\mu(1-\lambda)}{2}(3-3\alpha+\alpha^2)+\frac{\mu^2 \lambda}{2}(3-2\alpha)
    \end{equation*}
    \begin{equation*}
        \pi_4^d=\frac{\mu\lambda}{2},\quad \pi_6^d=\frac{\mu\lambda}{2}(1-\alpha),\quad  \pi_{10}^d=0
    \end{equation*}
    Thus, the reward of this policy would be:
    \begin{equation}
    \begin{aligned}
        &AR_b(\lambda,\mu)\\&
        =\mu(3-3\alpha+\alpha^2)+\lambda\mu (\mu(3-2\alpha)-(\alpha-1)^2)
        \\ &= AR_a (\lambda,\mu)+ \lambda\mu\left(-3\mu^2+(4\lambda-5)\mu+(\lambda-1)^2\right)
    \end{aligned}
    \end{equation}
     \item Only drops in the state ((2,1,0),1,0): For this policy, we have:
     \begin{equation*}
     \begin{aligned}
         \pi_2^d &=\frac{\mu}{2}(1-\lambda-\frac{\lambda(1-\lambda)(1-\mu)}{1-\lambda\mu})(3-3\alpha+\alpha^2)\\ & +\frac{\mu^2\lambda}{2}(1+\frac{(1-\lambda)(1-\mu)}{1-\lambda\mu})(3-2\alpha)
     \end{aligned}
    \end{equation*}
    \begin{equation*}
        \pi_4^d=\frac{\mu\lambda}{2}(1+\frac{(1-\lambda)(1-\mu)}{1-\lambda\mu})
    \end{equation*}
    \begin{equation*}
        \pi_6^d=\frac{\mu\lambda}{2}(1-\alpha),\quad  \pi_{10}^d=0
    \end{equation*}
    Let $K(\lambda,\mu)=\frac{(1-\lambda)(1-\mu)}{1-\lambda\mu}$. The reward will be:
    \begin{equation}
    \begin{aligned}
        &AR_c(\lambda,\mu)\\
        &=(\lambda,\mu)+\mu\lambda K(\lambda,\mu)\left(\mu (3-2\alpha)+1 -(3-3\alpha+\alpha^2)\right)\\
        &=AR_b(\lambda,\mu)-\mu\lambda K(\lambda,\mu)(3\mu^2-6\mu+\lambda^2-3\lambda+2) 
    \end{aligned}
    \end{equation}
    \item Drops in none of the above states: If we decide on not dropping packets in any of the nontrivial states, we would have:
    \begin{equation*}
     \begin{aligned}
         \pi_2^d &=\frac{\mu}{2}(1-\frac{\lambda(1-\lambda)(1-\mu)}{1-\lambda\mu})(3-3\alpha+\alpha^2)\\ & +\frac{\mu^2\lambda}{2}(1-\lambda+\frac{(1-\lambda)(1-\mu)}{1-\lambda\mu})(3-2\alpha)\\
         & + \frac{\mu^3\lambda^2}{2}
     \end{aligned}
    \end{equation*}
    \begin{equation*}
        \pi_4^d=\frac{\mu\lambda}{2}(1+\frac{(1-\lambda)(1-\mu)}{1-\lambda\mu})-\frac{\mu\lambda^2}{2}
    \end{equation*}
    \begin{equation*}
        \pi_6^d=\frac{\mu\lambda}{2}(1-\alpha)+\frac{\mu\lambda^2}{2}(\lambda+2\mu-1)
    \end{equation*}
    \begin{equation*}
          \pi_{10}^d=\frac{\mu\lambda^2}{2}
    \end{equation*}
    Let $K(\lambda,\mu)=\frac{(1-\lambda)(1-\mu)}{1-\lambda\mu}$. The reward will be:
    \begin{equation}
    \begin{aligned}
        &AR_d(\lambda,\mu)\\&=AR_c(\lambda,\mu)+\mu\lambda^2\left(3\mu^2+(2\lambda-1)\mu+\lambda-1\right) 
    \end{aligned}
    \end{equation}
\end{enumerate}
\subsection{Choice of optimal policy}  
 In this part, we compute the conditions for which any of these policies become optimal.
 \begin{itemize}
     \item Comparison between policies (a) and (b): Plugging in the results from the previous part, we get: 
     \begin{equation*}
     \begin{aligned}
         AR_b(\lambda,\mu) &\geq AR_a(\lambda,\mu)\\
         &\updownarrow \\
         \lambda\mu (-3\mu^2+(4\lambda&-5)\mu+(\lambda-1)^2) \geq 0\\
         &\updownarrow \\
        -3\mu^2+(4\lambda&-5)\mu+(\lambda-1)^2) \geq 0
     \end{aligned}
     \end{equation*}
    Solving the second-degree equation we get:
    \begin{equation}\label{bound(a)}
      \begin{aligned}
         AR_b(\lambda,\mu) &\geq AR_a(\lambda,\mu)\\
         &\updownarrow \\
        \mu &\geq \frac{5-4\lambda-\sqrt{4\lambda^2-16\lambda+13}}{6}
     \end{aligned}  
    \end{equation}
     Thus, policy (b) is better than policy (a) if and only if equation \ref{bound(a)} holds.
     \item Comparison between policies (b) and (c): Plugging in the results from the previous part, we get: 
     \begin{equation*}
     \begin{aligned}
         AR_c(\lambda,\mu) &\geq AR_b(\lambda,\mu)\\
         &\updownarrow \\
         \mu\lambda K(\lambda,\mu)(3\mu^2-6\mu&+\lambda^2-3\lambda+2) \leq 0\\
         &\updownarrow \\
       (3\mu^2-6\mu+\lambda^2&-3\lambda+2) \geq 0
     \end{aligned}
     \end{equation*}
    Solving the second-degree equation, we get:
    \begin{equation}\label{bound(b)}
      \begin{aligned}
         AR_c(\lambda,\mu) &\geq AR_b(\lambda,\mu)\\
         &\updownarrow \\
        \mu &\geq 1-\frac{\sqrt{3}}{3}\sqrt{-\lambda^2+3\lambda+1}
     \end{aligned}  
    \end{equation}
     Thus, policy (c) is better than policy (b) if and only if equation \ref{bound(b)} holds.
 \item Comparison between policies (c) and (d): Finally, for policies (c) and (d) we have 
     \begin{equation*}
     \begin{aligned}
         AR_d(\lambda,\mu) &\geq AR_c(\lambda,\mu)\\
         &\updownarrow \\
         \mu\lambda^2(3\mu^2+(2&\lambda-1)\mu+\lambda-1)  \geq 0\\
         &\updownarrow \\
       3\mu^2+(2\lambda&-1)\mu+\lambda-1 \geq 0
     \end{aligned}
     \end{equation*}
    Solving the second-degree equation, we get:
    \begin{equation}\label{bound(c)}
      \begin{aligned}
         AR_d(\lambda,\mu) &\geq AR_c(\lambda,\mu)\\
         &\updownarrow \\
        \mu &\geq \frac{1-2\lambda+\sqrt{4\lambda^2-16\lambda+13}}{6}
     \end{aligned}  
    \end{equation}
     Thus, policy (d) is better than policy (c) if and only if equation \ref{bound(c)} holds.
 \end{itemize}
 Fig. \ref{D3} shows the boundaries derived above. These boundaries do not cross, and thus, the analysis in section \ref{SectionD3} is done.
\end{document}